\documentclass[%        Class options:
aps,%                   American Physical Society
prd,%                   Physical Review D
showpacs,%              Displays PACS after abstract
preprint,%              Preprint layout
tightenlines,%          Single spaced lines
superscriptaddress,%    Authors' addresses linked with superscripts
nofootinbib,%           Does not treat footnotes as references
floatfix]%              Fixes float errors
{revtex4}%              REVTEX 4 Package used
\usepackage{graphicx,%  Default Latex 2eps package for embedding figures
}
\begin{document}
%
%===========================================================================
\title{Solar neutrino oscillation parameters\\ after first KamLAND results}
%===========================================================================
%
\author{        G.L.~Fogli}
\affiliation{   Dipartimento di Fisica
                and Sezione INFN di Bari\\
                Via Amendola 173, 70126 Bari, Italy\\}
\author{        E.~Lisi}
\affiliation{   Dipartimento di Fisica
                and Sezione INFN di Bari\\
                Via Amendola 173, 70126 Bari, Italy\\}
\author{        A.~Marrone}
\affiliation{   Dipartimento di Fisica
                and Sezione INFN di Bari\\
                Via Amendola 173, 70126 Bari, Italy\\}
\author{        D.~Montanino}
\affiliation{   Dipartimento di Scienza dei Materiali
                and Sezione INFN di Lecce\\
                Via Arnesano, 73100 Lecce, Italy\\}
\author{        A.~Palazzo}
\affiliation{   Dipartimento di Fisica
                and Sezione INFN di Bari\\
                Via Amendola 173, 70126 Bari, Italy\\}
\author{        A.M.~Rotunno}
\affiliation{   Dipartimento di Fisica
                and Sezione INFN di Bari\\
                Via Amendola 173, 70126 Bari, Italy\\}

\begin{abstract}%...........................................................

We analyze the energy spectrum of reactor neutrino events recently
observed in the Kamioka \hbox{Liquid} scintillator Anti-Neutrino
Detector (KamLAND) and combine them with solar and  terrestrial
neutrino data, in the context of two- and three-family active
neutrino oscillations. In the $2\nu$ case, we find that the
solution to the solar neutrino problem at large mixing angle (LMA)
is basically split into two sub-regions, that we denote as LMA-I
and LMA-II. The LMA-I solution, characterized by lower values of
the squared neutrino mass gap, is favored by the global data fit.
This picture is not significantly modified in the $3\nu$ mixing
case. A brief discussion is given about the discrimination of the
LMA-I and LMA-II solutions with future KamLAND data. In both the
$2\nu$ and $3\nu$ cases, we present a detailed analysis of the
post-KamLAND bounds on the  oscillation parameters.
\end{abstract}%.............................................................
\medskip
\pacs{%         PACS Numbers:
26.65.+t, 13.15.+g, 14.60.Pq, 91.35.-x} \maketitle

%%%%%%%%%%%%%%%%%%%%%%%%%%%%%%%%%%%%%%%%%%%%%%%%%%%%%%%%%%%%%%%%%%%%%%%%%%%%%%%%
\section{Introduction}
%%%%%%%%%%%%%%%%%%%%%%%%%%%%%%%%%%%%%%%%%%%%%%%%%%%%%%%%%%%%%%%%%%%%%%%%%%%%%%%%

The year 2002 is likely to be remembered as the {\em annus
mirabilis\/} of solar neutrino physics. On April 20, direct and
highly significant evidence for $\nu_e$ flavor change into active
states was announced by the Sudbury Neutrino Observatory (SNO)
experiment \cite{AhNC}, crowning a four-decade long \cite{Deca}
series of beautiful observations
\cite{Cl98,Ksol,Ab02,Ha99,Ki02,Fu01,Fu02,AhCC,AhDN} of the solar
$\nu_e$ flux deficit \cite{Ba89,BP00}. On October 8, the role of
solar neutrino physics in shaping modern science was recognized
through the Nobel Prize jointly awarded to Raymond Davis, Jr., and
Masatoshi Koshiba, for their pioneering contributions to the
detection of cosmic neutrinos \cite{Nobe}. Finally, on December 6,
clear ``terrestrial'' evidence for the oscillation solution to the
solar neutrino deficit was reported by the Kamioka Liquid
scintillator AntiNeutrino Detector (KamLAND), through the
observation of long-baseline reactor $\overline \nu_e$
disappearance \cite{Kam1}. The seminal idea of studying lepton
physics by detecting solar \cite{Po46,Al49} and reactor
\cite{Po46,Al49,Re53} neutrinos keeps thus bearing fruits after
more than 50 years.

The KamLAND observation of $\overline\nu_e$ disappearance
\cite{Kam1} confirms the current interpretation of solar neutrino
data \cite{AhDN,Fu02,AllS,GetM,Las3} in terms of
$\nu_e\to\nu_{\mu,\tau}$ oscillations induced by neutrino mass and
mixing \cite{Pont,Maki}, and restricts the corresponding parameter
space $(\delta m^2,\theta_{12})$ within the so-called large mixing
angle (LMA) region. In this region, globally favored by solar
neutrino data \cite{Smir}, matter effects \cite{Matt,Barg} in
adiabatic regime \cite{Adia,Matt} are expected to dominate the
dynamics of flavor transitions in the Sun (see, e.g.,
\cite{LasA}). The KamLAND spectral data appear to exclude some
significant portions of the LMA solution \cite{Kam1}, where the
predicted spectrum distortions
\cite{Mu02,Marf,Barb,Go01,Vi02,GoKL,Al02,Ho02,Go02,Las3} would be
in conflict with observations \cite{Kam1}.

In this paper we analyze the first KamLAND spectral data
\cite{Kam1} and combine them with current solar neutrino data
\cite{Cl98,Ab02,Ha99,Ki02,Fu02,AhNC,AhCC,AhDN}, assuming two- and
three-flavor oscillations of active neutrinos \cite{Las3}, in
order to determine the surviving sub-regions of the LMA solution.
In the analysis we include the CHOOZ reactor data \cite{CHOO} and,
in the $3\nu$ case, also the relevant constraints on the larger
mass gap $\Delta m^2\gg \delta m^2$, coming from the
Super-Kamiokande (SK) atmospheric \cite{Shio} and KEK-to-Kamioka
(K2K) accelerator \cite{K2Ke,K2Ks} neutrino experiments, according
to the approach developed in \cite{Las3}.%
%-------------------------------------------------------------
\footnote{Our notation for the squared neutrino mass spectrum is
$(m^2_1,m^2_2,m^2_3)=(-\delta m^2/2,+\delta m^2,\pm\Delta m^2)$ as
in \cite{Las3}, the sign of $\Delta m^2$ being associated to
normal $(+)$ or inverted $(-)$ hierarchy. Unless otherwise
noticed, normal hierarchy is assumed. The mixing angles
$\theta_{ij}$ are defined as in the quark case \cite{Las3,PDGr}.}
%------------------------------------------------------------

In the $2\nu$ case, we find that the inclusion of the KamLAND
spectrum basically splits the LMA solution into two sub-regions at
``small'' and ``large'' $\delta m^2$, which we call LMA-I and
LMA-II, respectively (the LMA-I solution being preferred by the
data). Such regions are only slightly modified in the presence of
$3\nu$ mixing, namely, for nonzero values of the mixing angle
$\theta_{13}$. We also present updated bounds in the $3\nu$
parameter space $(\delta m^2,\theta_{12},\theta_{13})$.

The structure of this paper is as follows. In Sec.~II we describe
our analysis of the KamLAND data. In Sec.~III we combine such
measurements with solar and CHOOZ data, assuming $2\nu$ mixing. In
Sec.~IV we extend the analysis to $3\nu$ mixing, including the
SK+K2K terrestrial neutrino constraints. We draw our conclusions
in Sec.~V.

\section{Kamland data input}

In our KamLAND data analysis, we use the absolute spectrum of
events reported in \cite{Kam1}, taken above a background-safe
analysis threshold of 2.6 MeV in visible energy $E$.%
%-------------------------------------------------
\footnote{The ``visible'' or ``prompt'' energy $E$ is defined as
$E=E_\nu-(M_n-M_p)+m_e\simeq E_\nu-0.8$~MeV.}
%-----------------------------------------------
The events below such threshold might contain a significant
component of geological $\overline\nu_e$'s \cite{GeoN}, whose
large normalization uncertainties are poorly constrained at
present by the KamLAND data themselves \cite{Kam1}. Above 2.6 MeV,
a total of 54 events is found (including at most one possible
background candidate), against 86.8 events expected from reactors
\cite{Kam1}.

The observed energy spectrum of events is analyzed as in
\cite{Las3}, with the following improvements. According to
\cite{Kam1}, we adopt: (1) Relative fuel components $^{235}$U :
$^{238}$U : $^{239}$Pu : $^{241}$Pu = 0.568 : 0.078 : 0.297 :
0.057; (2) Absolute normalization of 86.8 events for no
oscillations; (3) Energy resolution width equal to
$7.5\%/E^{1/2}$; (4) Thirteen bins in visible energy above 2.6
MeV, with 0.425 MeV width. The experimental and theoretical
(oscillated) number of events in each bin are denoted as
$N_i^\mathrm{expt}$ and $N_i^\mathrm{osc}$, respectively
$(i=1,\dots,13)$.

Concerning the systematic errors, the information in \cite{Kam1}
does not allow to trace the magnitude and the bin-to-bin
correlations of each component. We have then approximately grouped
such uncertainties into two main components, labelled by the index
$k=1,\, 2$: (1) energy scale error; and (2) overall normalization
error. The first (second) uncertainty basically shifts the
spectrum along the energy (event) coordinate. More precisely, the
effect of the energy scale error $(k=1)$ is estimated by taking
the (symmetrized) fractional differences $c^1_i$ between the
values of $N_i^\mathrm{osc}$ calculated with and without shifts of
$\pm 1.91\%$ \cite{Kam1} in the true visible energy, for each
point in the oscillation parameter space. The normalization
uncertainty $(k=2)$ is here assumed to collect all the other error
components listed in \cite{Kam1}, up to a total of about 6\%:
$c^2_i=0.06$. The systematic shifts $c^k_i$ of the
$N_i^\mathrm{osc}$'s are then implemented through the ``pull
approach'' described in \cite{GetM}, namely, through linear
deviations $N_i^\mathrm{osc}\to
N_i^\mathrm{theo}=N_i^\mathrm{osc}(1+\sum_k c^k_i \xi_k)$, where
the $\xi_k$'s are univariate Gaussian random variables. In this
way, systematic correlations are also accounted for \cite{GetM}.

Concerning the statistical errors, the presence of bins with few
or zero events requires  Poisson statistics, that we approximately
implement through the $\chi^2$-like recipe suggested in the Review
of Particle Properties \cite{PDGr}.  A quadratic penalty in the
pulls $\xi_k$ is introduced to account for the systematics
\cite{GetM,Lind}. The final $\chi^2$ function for KamLAND (KL) is
thus
%.................................................................
\begin{equation}\label{chi}
\chi^2_\mathrm{KL} = \min_{\{\xi_k\}} \left[  \sum_{i=1}^{13} 2
\left( N_i^\mathrm{theo}-N_i^\mathrm{exp}+N_i^\mathrm{exp}\ln
\frac{N_i^\mathrm{exp}}{N_i^\mathrm{theo}}\right)+\sum_{k=1,2}
\xi^2_k \right]\ ,
\end{equation}
%.................................................................
where the $i$-th logarithmic term is dropped if
$N_i^\mathrm{exp}=0$ \cite{PDGr}. By expanding the first sum in
the $\chi^2$ at first order in the  shifts $c^k_i$, the
minimization in Eq.~(\ref{chi}) becomes elementary.

Some final remarks are in order. A detailed comparison of the
above function $\chi^2_\mathrm{KL}$ with the one adopted by the
KamLAND collaboration is not currently possible, since the latter
is given in a symbolic form in \cite{Kam1}. However, one can
observe that: (1) The KamLAND collaboration splits the total rate
and spectrum shape information, while we prefer to use the
absolute spectrum; (2) The KamLAND collaboration can take into
account a (very small) background component above 2.6 MeV and four
sources of systematics while, for a lack of information, we
neglect such small background, and group the systematics into two
main sources. For such reasons, we do not expect perfect
quantitative agreement between the KamLAND official oscillation
analysis and ours.  In fact, we find that the 95\% C.L.\ contours
of the rate+shape analysis in Fig.~6 of \cite{Kam1} are accurately
reproduced by our Eq.~(\ref{chi}) at a slightly lower C.L.
($90\%$). This loss of statistical power appears tolerable to us,
in the light of the limited experimental information which is currently available.%
%----------------------------------------
\footnote{A better evaluation and decomposition of systematic
effects and an improved $\chi^2_\mathrm{KL}$ definition will be
possible when more detailed KamLAND information on spectrum shape
errors will become public.}
%-----------------------------------------

\section{$2\nu$ analysis}

The updated $2\nu$ analysis of current solar+CHOOZ neutrino data,
as performed in \cite{Las3}, is presented here for the sake of
completeness. The fit includes 81 solar neutrino observables
\cite{GetM,Las3} and 14 CHOOZ spectrum bins \cite{CHOO,Las3}, for
a total of 95 data points. The best-fit point and its
$\chi^2_{\mathrm{\min}}$ value are given in the first row of
Table~I. The $\Delta\chi^2$ expansion around the minimum, relevant
for the estimation of the oscillation parameters $(\delta
m^2,\sin^2\theta_{12})$, is shown in Fig.~1, where we have
restricted the $\delta m^2$ range to the only three decades
relevant for the LMA solution and for the following KamLAND
analysis. Notice that the scale is linear in the
$\sin^2\theta_{12}$ variable.

The $2\nu$ analysis of KamLAND is performed by using
Eq.~(\ref{chi}), which gives the absolute $\chi^2_\mathrm{min}$
value reported in the second row of Table~I. For later purposes,
we also quote the second best fit parameters. Expansion around the
absolute minimum gives the C.L.\ contours shown in Fig.~2.%
%----------------------------------------------------------------------
\footnote{No oscillations (basically, oscillations with $\delta
m^2\lesssim 10^{-6}$ eV$^2$) provide a very bad fit, $\Delta\chi^2
= 14.1$.}
%------------------------------------------------------------------
There appears to be a ``tower'' of solutions which tend to merge
and become indistinguishable for increasing $\delta m^2$; the
lower three ones are, however, rather well separated at 90\% C.L.
Notice that our allowed regions are slightly larger (i.e., less
constraining) than those in the rate+shape analysis of
\cite{Kam1}, as explained in Sec.~II.  One of the two
octant-symmetric best fits points in Fig.~2 (black dots) is
remarkably close to the best fit in Fig.~1 (see also Table~I). The
difference in location with respect to the KamLAND official
best-fit point at $(\delta
m^2/\mathrm{eV}^2,\sin^2\theta_{12})=(6.9\times 10^{-5},\,0.5)$
\cite{Kam1} is not statistically significant, amounting to a
variation $\Delta\chi^2\ll 1$.

%===========================================================================
%\begingroup
%\squeezetable
\begin{table}[t]
\caption{Results of the two-flavor oscillation analysis. The first
two columns report the data set used and the corresponding number
of data point $N_\mathrm{data}$, respectively. The third and
fourth columns give the position and $\chi^2_{\min}$ value of the
first and second best fit point (when applicable). For a
goodness-of-fit check, the $\chi^2_{\min}$ value should be
compared with its expected $\pm 1\sigma$ range, given by
$[N_\mathrm{DF} - \sqrt{2N_\mathrm{DF}},N_\mathrm{DF} +
\sqrt{2N_\mathrm{DF}}]$ \cite{PDGr}, where
$N_\mathrm{DF}=N_\mathrm{data}-2$ (see last column). All the
$\chi^2_\mathrm{min}$ values appear to be close to---or slightly
below---the lower end of such $\pm 1\sigma$ range, indicating that
the fits are somewhat better than statistically expected (although
not suspiciously so).}
\begin{ruledtabular}
\begin{tabular}{llcrc}
Experimental data set & No.\ of data & Best fit
point(s)\footnotemark[1] & $\chi^2_{\min}$ & $N_\mathrm{DF}\pm
\sqrt{2N_\mathrm{DF}}$ \\
\hline%---------------------------------------------------------------------------
Solar+CHOOZ         & 81+14     & (5.5,~0.305)   & 78.8  & [79.4,~106.6] \\
\hline%---------------------------------------------------------------------------
KamLAND             & 13        & (7.3,~0.335) &  6.1  & [6.3,~15.7]   \\
                    & 13        & (18.0,~0.270)
                                                 &  7.9  & [6.3,~15.7]   \\
\hline%---------------------------------------------------------------------------
Solar+CHOOZ+KamLAND & 81+14+13  & (7.3,~0.315)\footnotemark[2]& 85.2  & [91.4,~120.6] \\
                    & 81+14+13  & (15.4,~0.300)\footnotemark[3]
                                                 & 90.6  & [91.4,~120.6]
%================================================================================
\end{tabular}
\end{ruledtabular}
\footnotetext[1]{Coordinates are ($\delta m^2/10^{-5}$
eV$^2$,~$\sin^2\theta_{12}$). For KamLAND only,
$\sin^2\theta_{12}$ and $1-\sin^2\theta_{12}$ are equivalent.}
\footnotetext[2]{Global best fit (LMA-I).} \footnotetext[3]{Second
global best fit (LMA-II).}
\end{table}
%\endgroup
%==================================================================================

The combination of the solar+CHOOZ results in Fig.~1 with the
KamLAND results in Fig.~2 gives the global $2\nu$ results shown in
Fig.~3, which represent the main result of this work. Two rather
distinct solutions, that we label LMA-I and LMA-II, are seen to
emerge. They are basically located at the intersection of the LMA
solution in Fig.~1 with two of the well-separated KamLAND
solutions in Fig.~2, and are characterized by the mass-mixing
parameters and $\chi^2$ values given in the last two rows of
Table~1. The LMA-I solution is clearly preferred by the data,
being close to the best fit points of both solar+CHOOZ and KamLAND
data. The LMA-II solution is located at a $\delta m^2$ value about
twice as large as for the LMA-I, but is separated from the latter
by a modest $\Delta\chi^2=5.4$ difference (dominated by solar
neutrino data). Indeed, if we conservatively demand a 99.73\%
C.L.\ for the allowed regions, the LMA-I and LMA-II solutions
appear to merge (and extend towards $\delta m^2\sim 3\times
10^{-4}$ eV$^2$) in a single broad solution. In any case, at any
chosen C.L., the allowed regions of Fig.~3 are significantly
smaller than those in Fig.~1. Therefore, with just 54 initial
events, the KamLAND experiment is not only able to select the LMA
region as {\em the\/} solution to the solar neutrino problem, but
can also significantly restrict the corresponding oscillation
parameter space. With several hundred events expected in the
forthcoming years, there are thus very good prospects to refine
the $(\delta m^2,\sin^2\theta_{12})$ parameter estimate
\cite{Las3}

The most important task for the next future appears to be the
confirmation of one of the two solutions in Fig.~3 through higher
statistics and, possibly, lower analysis threshold. Figure~4 shows
the absolute KamLAND energy spectra of reactor events predicted at
the LMA-I and LMA-II global best-fit points (last two rows of
Table~I), together with the no-oscillation spectrum, with the
current event normalization. The KamLAND data (with statistical
errors only) are also superposed above the analysis threshold (2.6
MeV). It can be seen that the main difference between the two
oscillated spectra is the position of the spectrum peak, which is
roughly aligned with the no-oscillation position for the LMA-II
case, while it is shifted at higher energies for the LMA-I case.
This feature might be a useful discrimination tool in the next
future. From Fig.~4, it appears also that the largest difference
between the LMA-I and LMA-II spectra in KamLAND occurs just in the
first bin below the current analysis threshold. Therefore, a
better understanding of the background below 2.6 MeV will be
highly beneficial. In any case, an increase in statistics by a
factor of a few appears necessary to disentangle the two
oscillated spectra in Fig.~4.

\section{$3\nu$ analysis}

The updated $3\nu$ analysis of solar+CHOOZ neutrino data,
including the constraints on the large (``atmospheric'') squared
mass gap $\Delta m^2$ coming from SK atmospheric \cite{Shio} and
K2K accelerator neutrino data \cite{K2Ke,K2Ks}, as performed in
\cite{Las3},  is
reported here for the sake of completeness.%
%------------------------------------------------------------------
\footnote{In \cite{Las3}, the combination of SK and K2K
constraints on the $\Delta m^2$ parameter was obtained by adding
the preliminary $\chi^2$ functions from the two collaborations, as
graphically presented in \cite{Shio,K2Ke}. We anticipate that such
combination appears to be in good agreement with a more refined
and joint reanalysis of SK atmospheric and K2K data \cite{Prog}.}
%------------------------------------------------------------------
Figure~5 shows the pre-KamLAND $3\nu$ results \cite{Las3} in the
parameter space $(\delta m^2,\sin^2\theta_{12},\sin^2\theta_{13})$
relevant for the LMA solution, shown through sections at four
different
values of $s^2_{13}=\sin^2\theta_{13}$.%
%----------------------------------------------------------------
\footnote{We remind that, in Fig.~5 and in the following $3\nu$
figures, the information on $\Delta m^2$ is projected away by
taking $\min_{\Delta m^2}(\chi^2)$, see \cite{Las3}.}
%------------------------------------------------------------------
For later purposes, we only notice that the upper bound on $\delta
m^2$ becomes stronger for higher $\sin^2\theta_{13}$, as a result
of the CHOOZ constraints. For a discussion of such $(\delta
m^2,\sin^2\theta_{13})$ anti-correlation, and for the pre-KamLAND
solutions below the LMA, see \cite{Las3} and references therein.

Figure~6 shows the results of our $3\nu$ fit to the KamLAND data,
in the same format as in Fig.~5. The allowed $3\nu$ regions appear
to get slightly enlarged (especially in the mixing parameter
$\theta_{12}$) for increasing values of $\sin^2\theta_{13}$,  as
expected \cite{GoKL}. In fact, for $\theta_{13}\neq 0$, part of
the KamLAND event disappearance is explained by averaged
oscillations driven by $\Delta m^2$ in the $(\nu_e,\nu_3)$ sector,
so that the overall oscillation amplitude $\sin^2 2\theta_{12}$ in
the $(\nu_1,\nu_2)$ sector is allowed to reach smaller values, and
the $\sin^2\theta_{12}$ range is correspondingly enlarged
\cite{GoKL}. The absolute $\chi^2$ minimum in Fig.~6 is reached
for $\theta_{13}\neq 0$, but such preference is not statistically
significant, as expected \cite{Marf}; indeed, we find a mere
$\Delta\chi^2=0.4$ increase of the KamLAND $\chi^2_{\min}$ value
for fixed $\sin^2\theta_{13}=0.06$. Concerning the variations of
the $(\delta m^2,\sin^2\theta_{12})$ best-fit coordinates for
fixed values of $\sin^2\theta_{13}$, we find that the $\delta m^2$
value is stable (and equal to $7.3\times 10^{-5}$ eV$^2$, as in
the $2\nu$ case), while $\sin^2\theta_{12}$ decreases from 0.335
to 0.225 when increasing $\sin^2\theta_{13}$ from 0 to 0.06, in
good qualitative agreement with the expectations discussed in
\cite{GoKL}.

However, the additional spread in $\sin^2\theta_{12}$ induced by
nonzero $\theta_{13}$ currently does not play any relevant role
when  KamLAND is combined with world neutrino data, for at least
two reasons: (1)  Pre-KamLAND bounds from solar+terrestrial data
currently dominate the constraints on $\sin^2\theta_{12}$, as
evident from a comparison of Fig.~5 and Fig.~6; (2) The likelihood
of genuine $3\nu$-induced effects rapidly decreases with
increasing $\theta_{13}$, because of the strong upper bounds on
such mixing parameter \cite{Las3}. Therefore, we do not expect any
significant enlargement of the $\sin^2\theta_{12}$ allowed range
from the pre- to the post-KamLAND $3\nu$ analysis, despite the
presence of such effects in KamLAND alone.

It should be noted that, strictly speaking, Fig.~6 is not exactly
a ``KamLAND only'' analysis, since we have implicitly taken some
pieces of information from terrestrial neutrino data. In
particular, we have implicitly assumed in Fig.~6 that: (i) the
``atmospheric'' squared mass splitting $\Delta m^2$ is
sufficiently high to be unresolved in KamLAND, and (2) the
relevant values of $\sin^2\theta_{13}$ are limited in the few
percent range. It is interesting to study the effect of an
explicit combination of such (atmospheric~+~CHOOZ) information
with KamLAND data. The results are given in Fig.~7. This figure
show that purely {\em terrestrial\/} neutrino data from
atmospheric (SK), accelerator (K2K) and reactor (KamLAND~+~CHOOZ)
neutrino experiments, by themselves, are now able to put both
upper and lower bounds on the {\em solar\/} parameters $(\delta
m^2,\sin^2\theta_{12})$. As previously noted in the comment to
Fig.~5, the CHOOZ upper bound on $\delta m^2$ becomes
stronger when $\sin^2\theta_{13}$ increases.%
%------------------------------------------------------------------
\footnote{Analogously,  the upper bound on $\sin^2\theta_{13}$
becomes stronger for increasing $\delta m^2$ \cite{BiCo}.}
%----------------------------------------------------------------------
The slight octant asymmetry  for $\sin^2\theta_{13}>0$ is due to
the fact the corresponding $3\nu$ CHOOZ probability is not
invariant under the change $\theta_{12}\to\pi/2-\theta_{12}$ for
fixed hierarchy \cite{QEIP} (assumed to be normal in Fig.~6). The
octant differences would be swapped by inverting the hierarchy
\cite{QEIP}. In practice, however, such octant asymmetries are
numerically irrelevant in the global fit which we now discuss.

Figure~8 shows the final $3\nu$ combination of  world
(solar+terrestrial) neutrino constraints, including  pre-KamLAND
data (Fig.~5) and the first KamLAND data (Fig.~6). For
\hbox{$\sin^2\theta_{13}=0$} (absolute best fit), we get the same
LMA-I and LMA-II solutions reported in Fig.~3 and in Table~I,
modulo the expected widening induced by one additional degree of
freedom in the $\Delta\chi^2$'s associated to each C.L.\ contour.
This widening gives, at 99\% C.L., marginal allowance for a third
solution with $\delta m^2\sim 2.5$--$3.2 \times 10^{-4}$ eV$^2$,
which we call
LMA-III.%
%---------------------------------------------------------
\footnote{ Notice that the LMA-I and LMA-III solutions differ by
$\Delta\chi^2\simeq 11$.}
%-------------------------------------------------------------
All solutions rapidly shrink for increasing $\sin^2\theta_{13}$,
the LMA-I being the most stable and the last to disappear. The
hierarchy is assumed to be normal in Fig.~8; the differences with
respect to the inverted hierarchy case (not shown) are completely
negligible.

Figure~9 represents a compendium of our post-KamLAND $3\nu$
analysis of all data (solar+terrestrial), in terms of the
$\Delta\chi^2$ function for each of the three variables $(\delta
m^2,\sin^2\theta_{12},\sin^2\theta_{13})$, the others being
projected (minimized) away. This figure should be compared with
the pre-KamLAND one in \cite{Las3}. The comparison shows that the
bounds on $\sin^2\theta_{13}$ remain basically unaltered in the
post-KamLAND era, while the mixing parameter $\sin^2\theta_{12}$
is clearly nailed down around the value $0.3$. The absolute and
second best minima (LMA-I and LMA-II) are also evident in the left
panel. A third minimum (LMA-III) at higher $\delta m^2$ is only
marginally allowed.

If we focus on the best-fit (LMA-I) solution, and adapt parabolic
functions around the absolute $\chi^2$ minima in terms of $\delta
m^2$ (in linear scale), $\sin^2\theta_{13}$, and
$\sin^2\theta_{13}$, we obtain from the post-KamLAND global
analysis the following approximate $\pm 1\sigma$ estimates
($\Delta \chi^2\simeq 1$) for the relevant solar $3\nu$
oscillation parameters,
%..............................................................
\begin{equation}\label{val}
\textrm{LMA-I\ }(\sim 1\sigma):\left\{\begin{array}{l}
\delta m^2\simeq (7.3\pm 0.8)\times 10^{-5} \mathrm{\ eV}^2\ ,\\
\sin^2\theta_{12}\simeq 0.315 \pm 0.035\ ,\\
\sin^2\theta_{13}\lesssim 0.017\ .
\end{array}\right.
\end{equation}
%..............................................................
The above ranges are meant to show that we are not far from a 10\%
determination (at $1\sigma$) of both $\delta m^2$ and
$\sin^2\theta_{12}$, but should not be quoted as a ``summary'' of
the current $3\nu$ situation. The technically correct reference
summary for our analysis is represented by the $\chi^2$ functions
of Fig.~9, which include the possibility of a second global best
fit (LMA-II), and maybe of a third best fit (LMA-III) in $\delta
m^2$. In particular, we stress that the LMA-II solution is
currently quite acceptable from a statistical point of view, and
that the history of  the solar neutrino problem teaches us that
one should not exclude a priori what may appear as the second
best-fit at a particular time.

Similar cautionary remarks also apply to the use of the initial
KamLAND data to refine the emerging indications of adiabatic
matter effects in the Sun \cite{Adia,Matt} through the approach
developed in \cite{LasA}. As stressed in \cite{LasA}, such
analysis will become sufficiently stable and constraining when a
single solution (either the LMA-I or the LMA-II) will be
(hopefully) clearly selected by KamLAND. In the meantime, further
SNO data are also expected to increase our confidence in the
occurrence of solar matter effects, see \cite{LasA}.

We conclude our $3\nu$ analysis by observing that, from the point
of view of future (very) long baseline (LBL) experiments, the
``solar'' neutrino parameters $\delta m^2$ and $\theta_{12}$ enter
the subleading oscillation amplitudes of both the ``golden''
channel $\nu_e\to\nu_\mu$ \cite{Gold} and the ``silver'' channel
$\nu_e\to\nu_\tau$ \cite{Silv} only through the combination
$\kappa=\delta m^2\sin 2\theta_{12}$, up to $O(\kappa^2)$ terms
included \cite{Gold,Silv}. Therefore, we think it useful to show
in Fig.~10 the global (solar+terrestrial) bounds on this specific
combination of parameters (in units of $10^{-5}$ eV$^2$ and in
linear scale), all other $3\nu$ variables being projected away.
The lower and higher minima in Fig.~10 correspond to the LMA-I and
LMA-II cases, respectively. By adapting a parabola in the
$\pm2\sigma$ region around each of the two minima, we get the
following $1\sigma$ ($\Delta \chi^2=1$) estimates for
$\kappa=\delta m^2\sin2\theta_{12}$: $\kappa\simeq 6.7\pm 0.7$
(LMA-I) and $\kappa\simeq 14.7\pm 1.7$ (LMA-II), in units of
$10^{-5}$ eV$^2$. In both cases, $\kappa$ appears to be already
constrained with a $\sim 11\%$ accuracy. These approximate
estimates---or, better, the accurate bounds in Fig.~10---may prove
useful in prospective studies of $\kappa$-dependent subleading
oscillation effects in LBL projects.

\section{Summary and prospects}

The KamLAND experiment has clearly selected the LMA region as the
solution to the solar neutrino problem, and has further reduced
the $(\delta m^2,\sin^2\theta_{12})$ parameter space for active
neutrino oscillations. In this work, we have studied such
parameter reduction in the the context of $2\nu$ and $3\nu$
oscillations, by using the limited KamLAND experimental
information which is publicly available. In the $2\nu$ case, we
find that the post-KamLAND LMA solution appears to be basically
split into two sub-regions, LMA-I and LMA-II. The LMA-I solution,
characterized by $\delta m^2\sim 7\times 10^{-5}$ eV$^2$ and
$\sin^2\theta_{12}\sim 0.3$, is preferred by the global fit. The
LMA-II solution represents the second best fit, at about twice the
value of $\delta m^2$ (see Table~I). This situation is not
significantly changed in the $3\nu$ case, for which we present a
global post-KamLAND analysis of solar and terrestrial data in the
$(\delta m^2,\sin^2\theta_{12},\sin^2\theta_{13})$ parameter
space. There are good prospects to separate the LMA-I and LMA-II
cases with future, higher-statistics KamLAND data, by looking at
the peak of the energy spectrum, and by lowering the current
analysis threshold (2.6 MeV) by at least $\sim 0.5$ MeV.

%%%%%%%%%%%%%%%%%%%%%%%%%%%%%%%%%%%%%%%%%%%%%%%%%%%%%%%%%%%%%%%%%%%%%%%%%%
\acknowledgments

This work was supported in part by INFN and in part by the Italian
{\em Ministero dell'Istruzione, Universit\`a e Ricerca\/} through
the ``Astroparticle Physics'' research project. We are grateful to
G.\ Gratta for prompt information about the release of the first
KamLAND results.

%%%%%%%%%%%%%%%%%%%%%%%%%%%%%%%%%%%%%%%%%%%%%%%%%%%%%%%%%%%%%%%%%%%%%%%%

%%%%%%%%%%%%%%%%%%%%%%%%%%%%%%%%%%%%%%%%%%%%%%%%%%%%%%%%%%%%%%%%%%%%%%%%

%---------------------------------------------------------------------------
\begin{figure}
\vspace*{0.6cm}\hspace*{-5mm}
\includegraphics[scale=0.85, bb= 100 100 500 720]{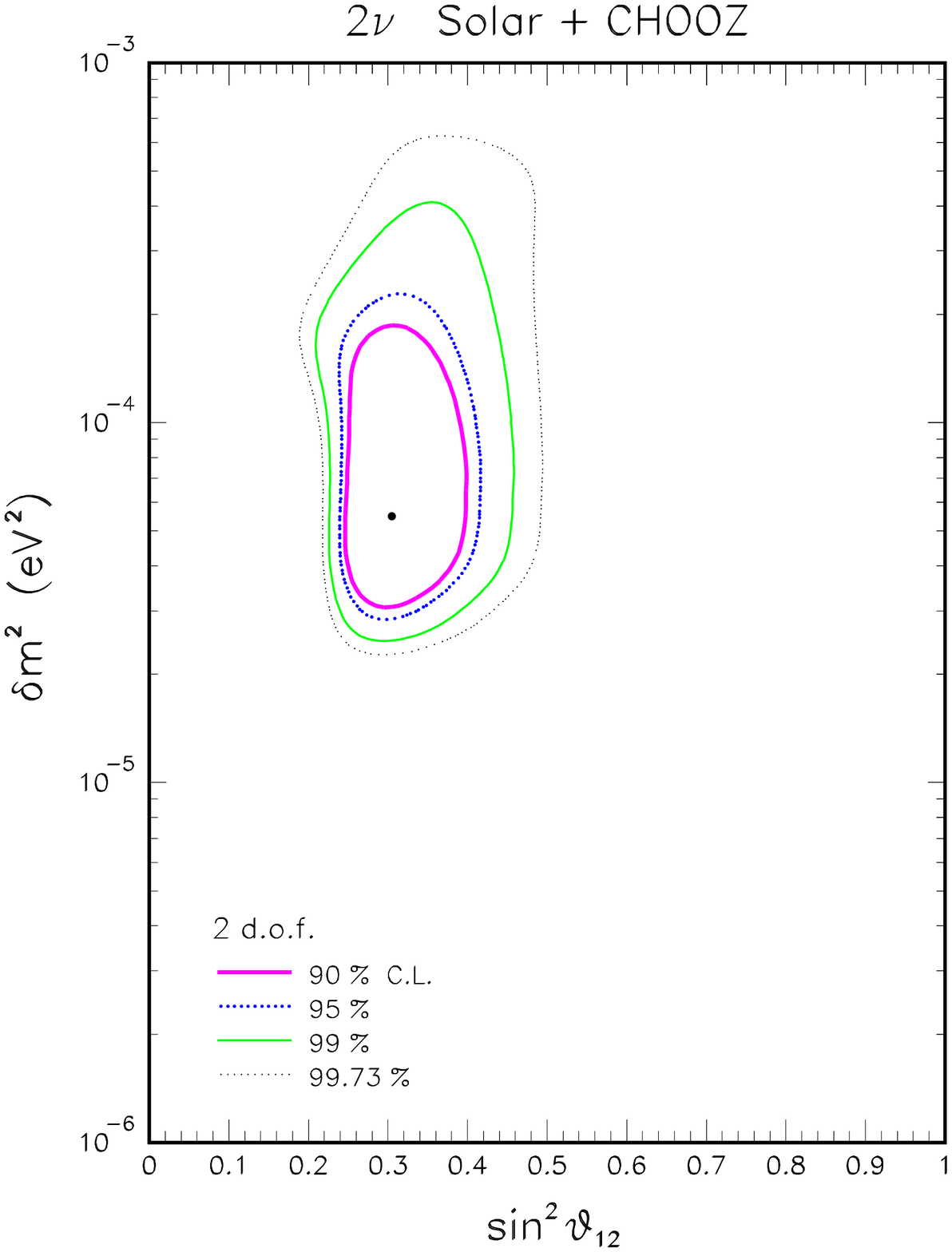}
\vspace*{+0.4cm} \caption{\label{fig1} Two-flavor active neutrino
oscillations: Global analysis of solar and CHOOZ neutrino data in
the $(\delta m^2,\sin^2\theta_{12})$ parameter space, restricted
to the LMA region. The best fit is indicated by a black dot. The
90, 95, 99, and 99.73\% C.L.\ contours correspond to $\Delta
\chi^2=4.61$, 5.99, 9.21, and 11.83, respectively (two degrees of
freedom). For solutions at lower $\delta m^2$---now ruled out by
KamLAND---see \cite{Las3}.}
\end{figure}

%---------------------------------------------------------------------------
\begin{figure}
\vspace*{0.6cm}\hspace*{-5mm}
\includegraphics[scale=0.85, bb= 100 100 500 720]{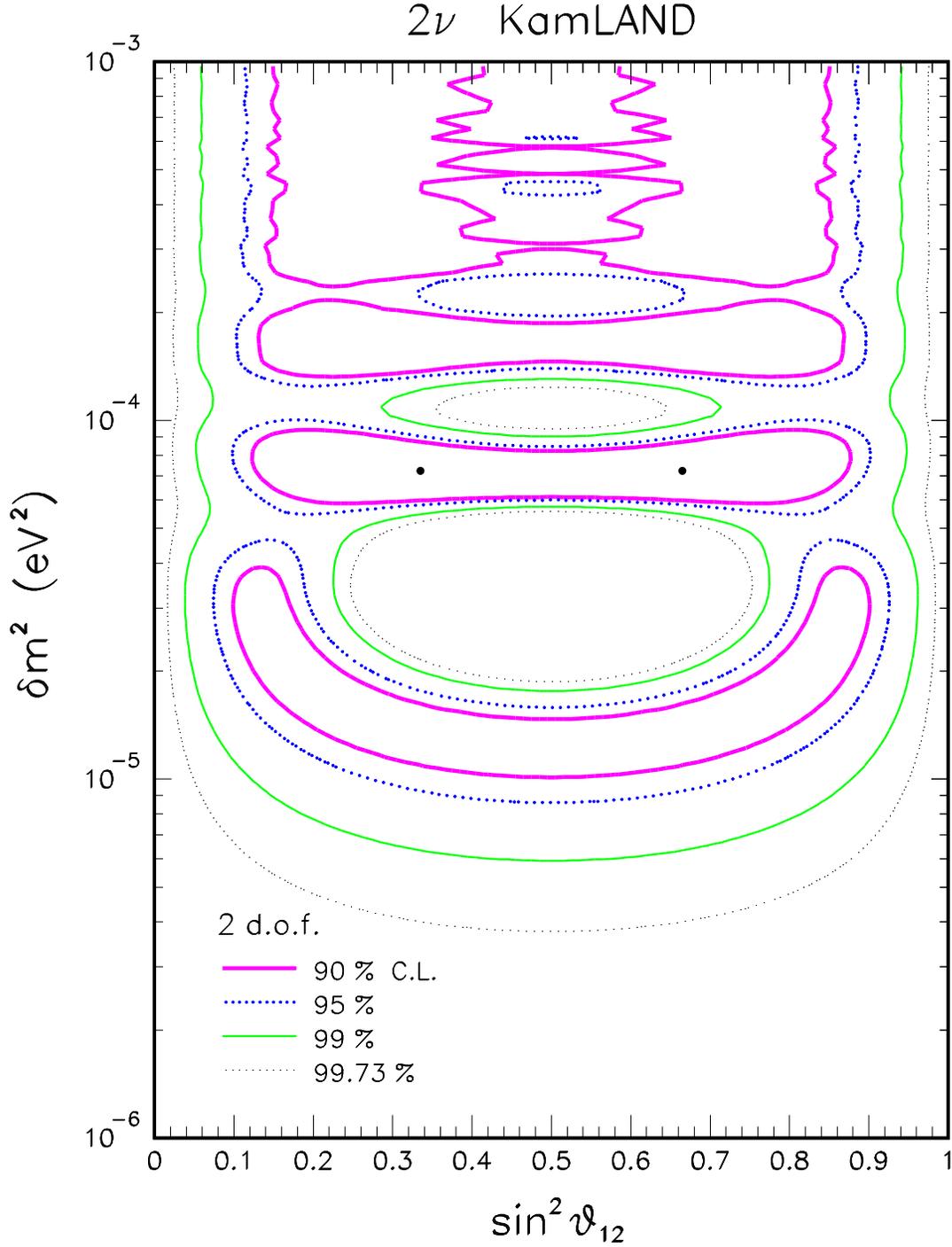}
\vspace*{+0.4cm} \caption{\label{fig2} Two-flavor active neutrino
oscillations: Analysis of the KamLAND energy spectrum data above
2.6 MeV in the $(\delta m^2,\sin^2\theta_{12})$ parameter space. A
``tower'' of octant-symmetric regions is allowed at different
values of $\delta m^2$. The three lower regions with $\delta
m^2\lesssim 2\times 10^{-4}$ eV$^2$ are well separated at 90\%
C.L., while the upper ones tend to merge in a continuum. The
symmetric best fits are indicated by black dots. The left dot is
remarkably close to the solar best fit in Fig.~1.}
\end{figure}

%---------------------------------------------------------------------------
\begin{figure}
\vspace*{0.6cm}\hspace*{-5mm}
\includegraphics[scale=0.85, bb= 100 100 500 720]{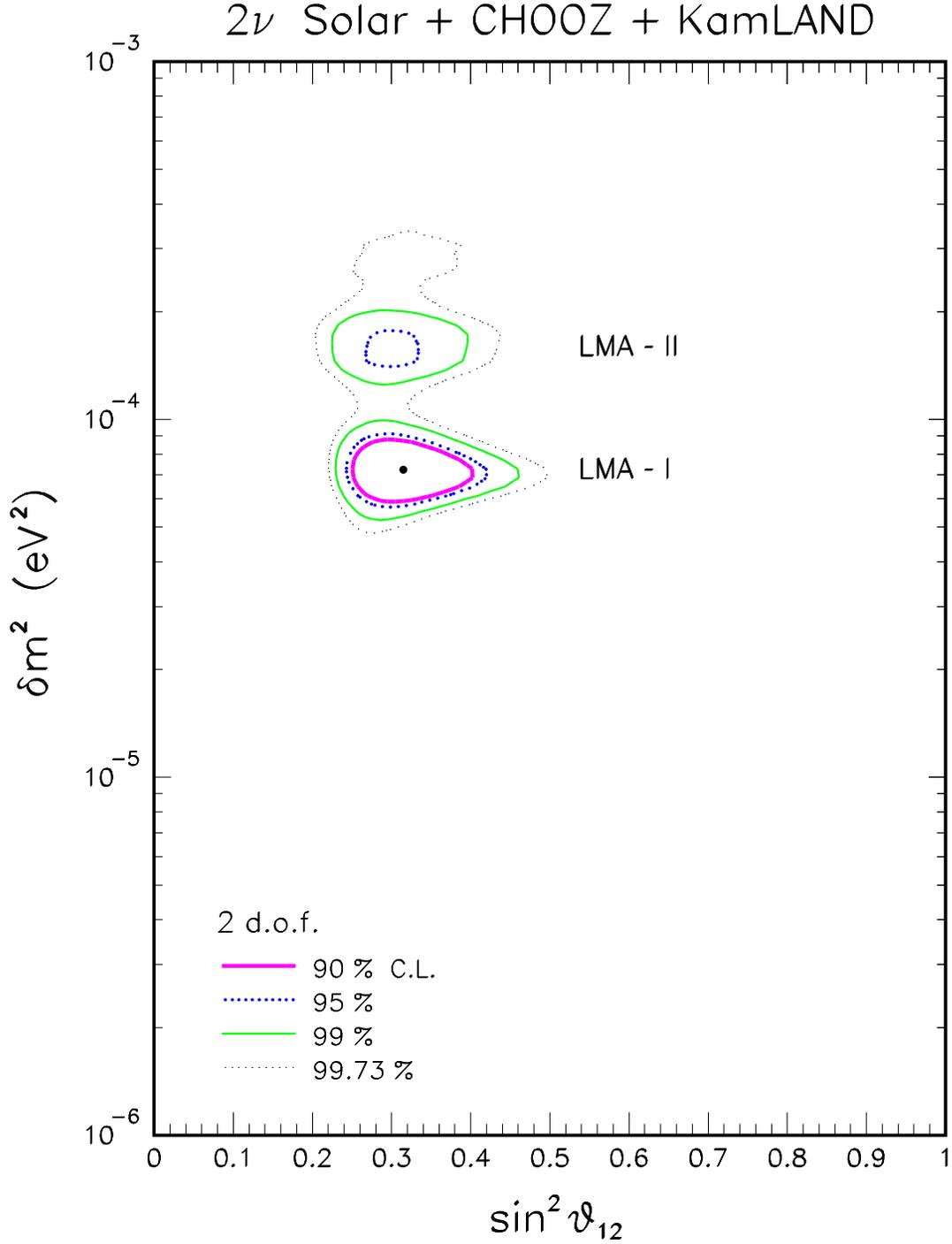}
\vspace*{+0.4cm} \caption{\label{fig3} Two-flavor active neutrino
oscillations: Global analysis of solar, CHOOZ, and KamLAND
neutrino data in the $(\delta m^2,\sin^2\theta_{12})$ parameter
space. With respect to Fig.~1, the LMA region is significantly
restricted, and appears to be split into two sub-regions (LMA-I
and LMA-II), well-separated at 99\% C.L. The best fit (black dot)
is found in the LMA-I solution. At 99.73\% C.L., the two solutions
merge into a single one, which  slightly extends towards $\delta
m^2\sim 3\times 10^{-4}$~eV$^2$.}
\end{figure}

%---------------------------------------------------------------------------
\begin{figure}
\vspace*{0.6cm}\hspace*{-5mm}
\includegraphics[scale=0.85, bb= 100 100 500 720]{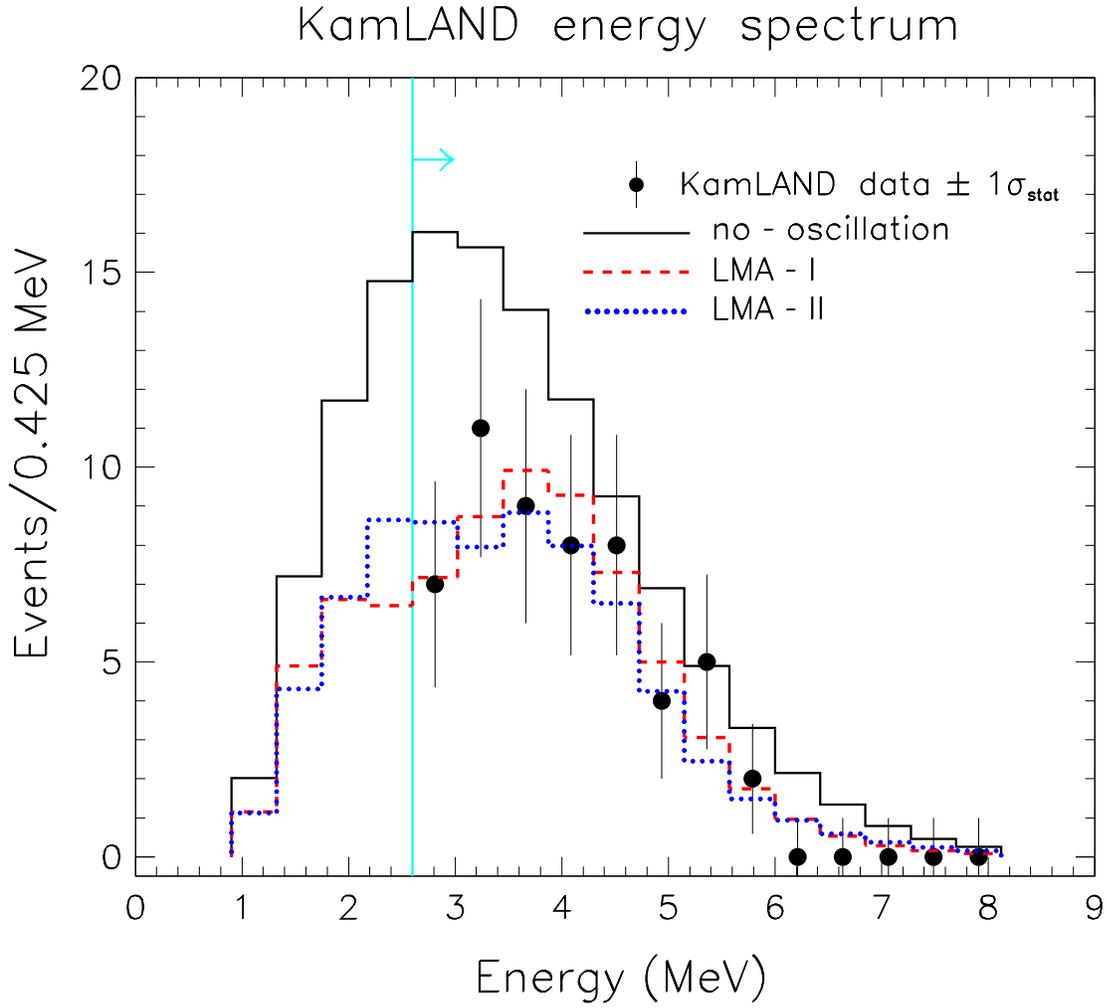}
\vspace*{-1.5cm} \caption{\label{fig4} Energy spectra of events in
KamLAND for the LMA-I and LMA-II global fit points in Table~II,
together with the no-oscillation spectrum. The current KamLAND
data are superposed  above the analysis threshold (2.6 MeV).}
\end{figure}

%---------------------------------------------------------------------------
\begin{figure}
\vspace*{0.6cm}\hspace*{-8mm}
\includegraphics[scale=0.9, bb= 100 100 500 720]{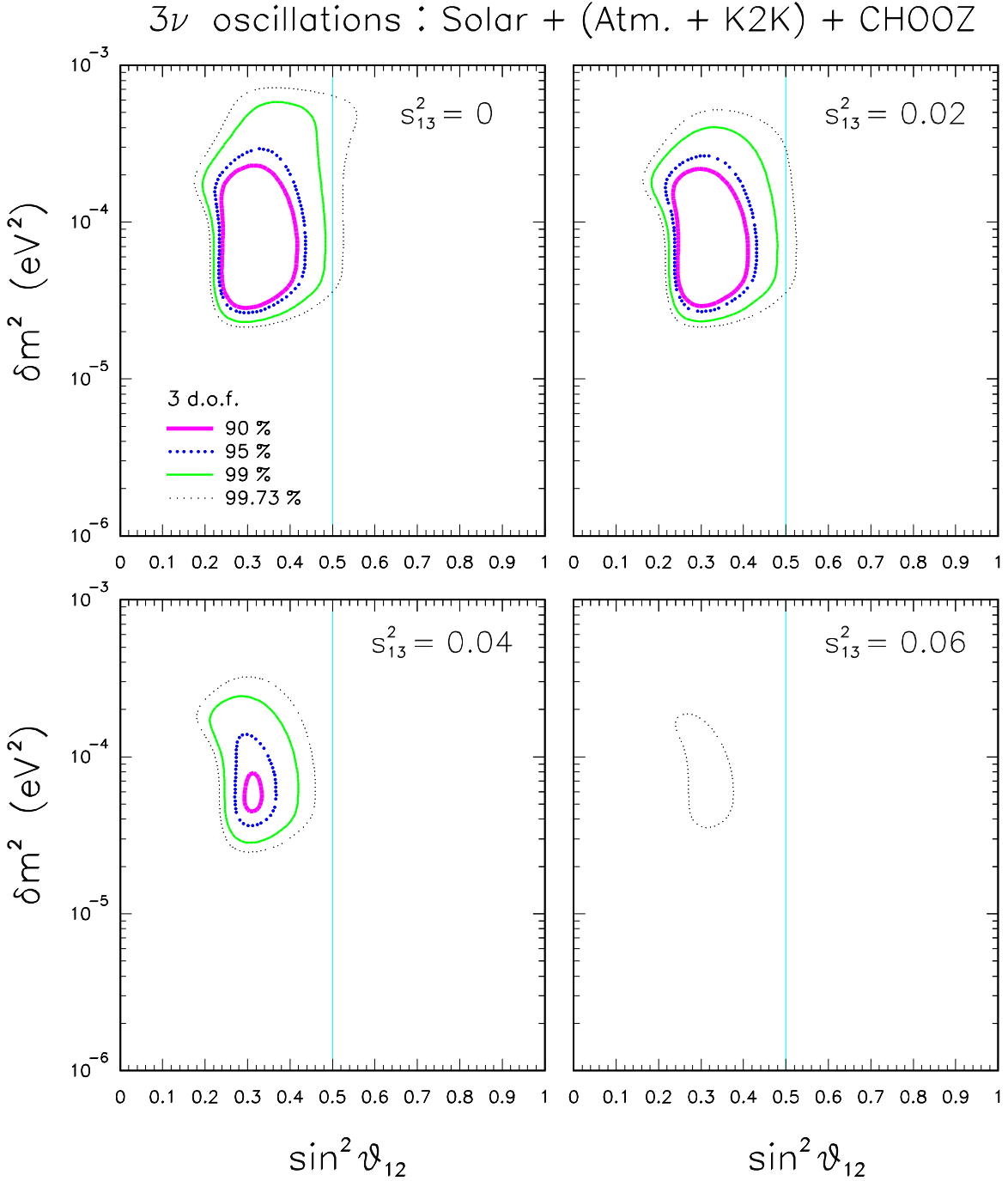}
\vspace*{-0.5cm} \caption{\label{fig5} Three-flavor active
neutrino oscillations: Global analysis of solar, CHOOZ, and
atmospheric+K2K neutrino data, shown as sections of the allowed
$(\delta m^2,\sin^2\theta_{12},\sin^2\theta_{13})$ parameter space
at four representative values of $s^2_{13}=\sin^2\theta_{13}$. The
best fit is reached in the $2\nu$ limit ($s^2_{13}=0$), while for
increasing $s^2_{13}$ the LMA solution shrinks and eventually
disappears. The 90, 95, 99, and 99.73\% C.L.\ contours correspond
to $\Delta \chi^2=6.25$, 7.82, 11.34, and 14.15, respectively
(three degrees of freedom). The ``atmospheric'' mass splitting
$\Delta m^2$ is minimized away in the fit (see also \cite{Las3}).}
\end{figure}

%---------------------------------------------------------------------------
\begin{figure}
\vspace*{0.6cm}\hspace*{-8mm}
\includegraphics[scale=0.9, bb= 100 100 500 720]{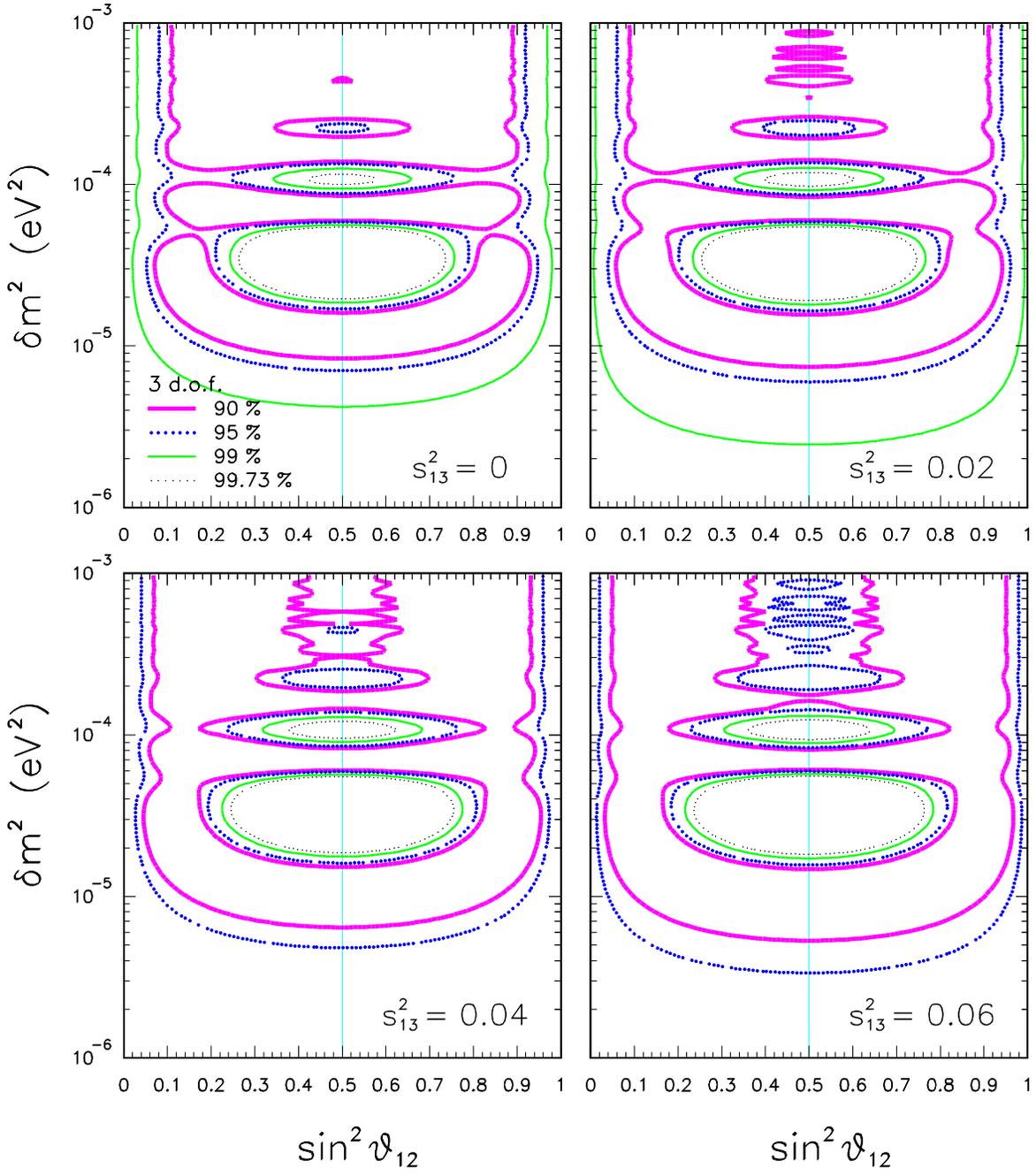}
\vspace*{-0.5cm} \caption{\label{fig6} Three-flavor active
neutrino oscillations: Global analysis of KamLAND data, shown in
the same format as Fig.~5. Notice the enlargement of the allowed
regions for increasing $s^2_{13}$.}
\end{figure}

%---------------------------------------------------------------------------
\begin{figure}
\vspace*{0.6cm}\hspace*{-8mm}
\includegraphics[scale=0.9, bb= 100 100 500 720]{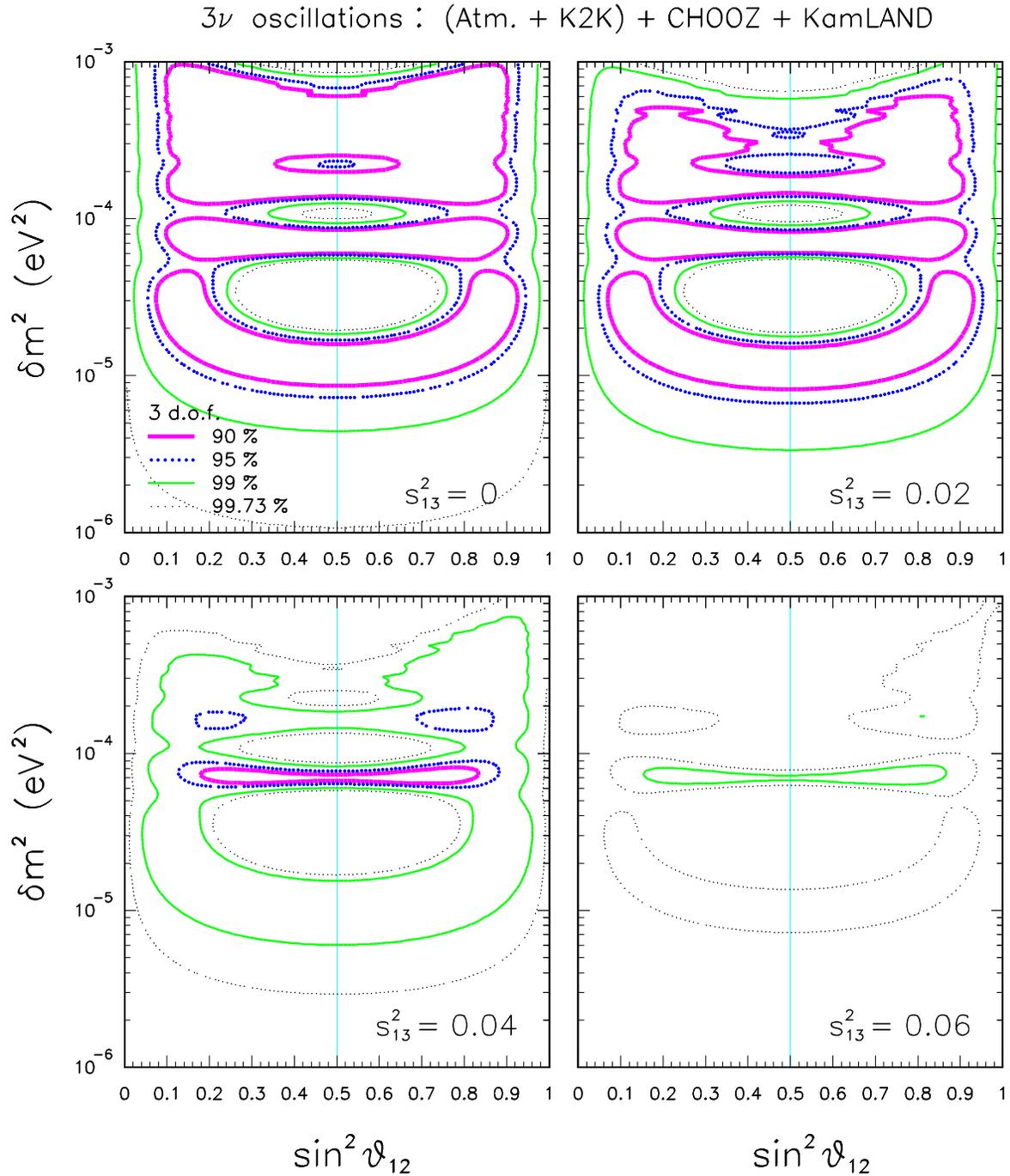}
\vspace*{-0.5cm} \caption{\label{fig7} Three-flavor active
neutrino oscillations: Global analysis of terrestrial data only
(including KamLAND), shown in the same format as Fig.~5. Notice
that {\em terrestrial} $\nu$ data alone can now put both upper and
lower bounds on the {\em solar} $\nu$ parameters ($\delta
m^2,\sin^2\theta_{12}$). See the text for details.}
\end{figure}

%---------------------------------------------------------------------------
\begin{figure}
\vspace*{0.6cm}\hspace*{-8mm}
\includegraphics[scale=0.9, bb= 100 100 500 720]{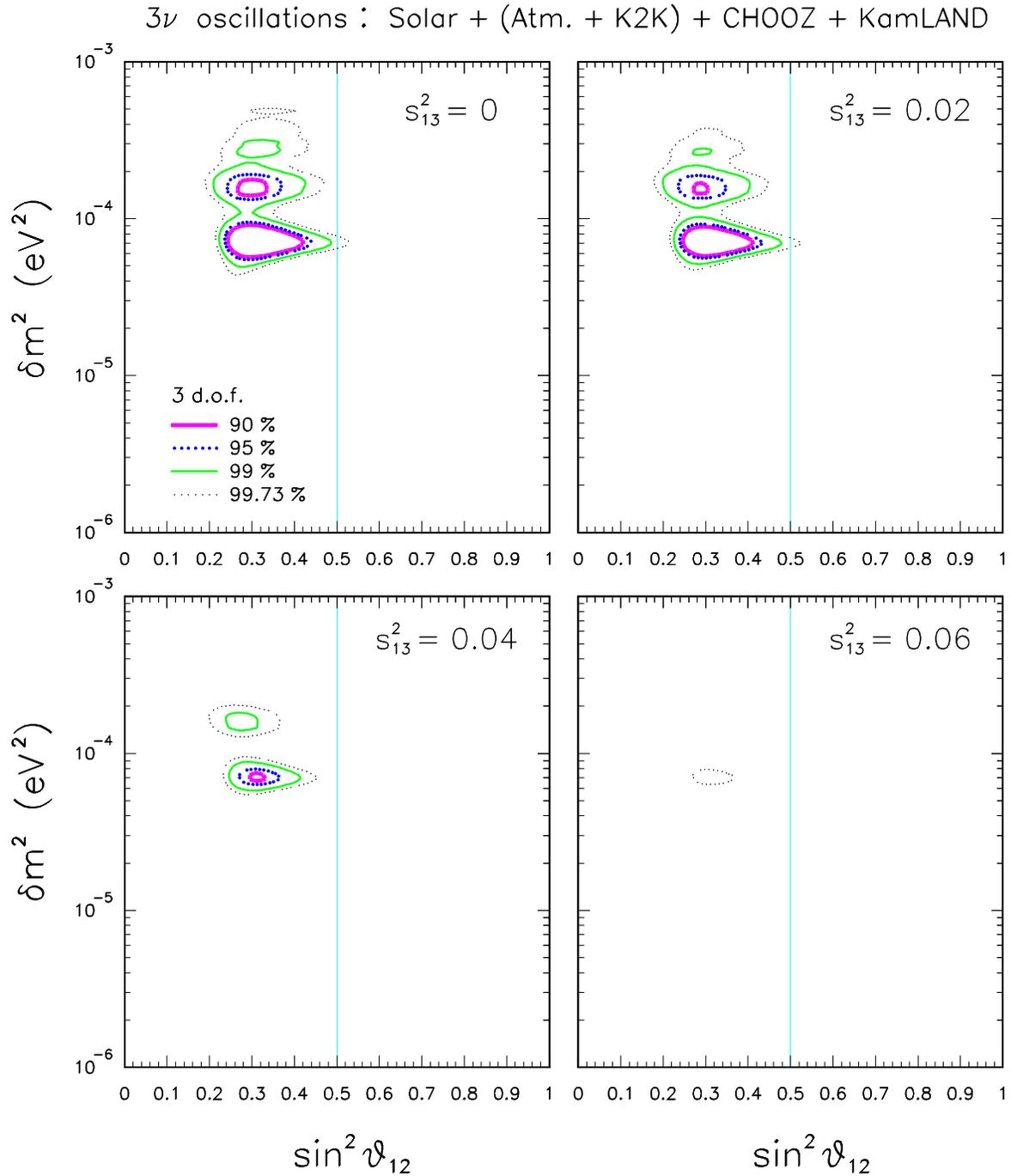}
\vspace*{-0.5cm} \caption{\label{fig8} Three-flavor active
neutrino oscillations: Global analysis of solar and terrestrial
data (including KamLAND), shown in the same format as Fig.~5. With
respect to the $2\nu$ case in Fig.~3, the higher $\Delta\chi^2$
tolerance induced by an extra degree of freedom (for any chosen
C.L.) marginally allows a third ``LMA-III'' solution at 99\% C.L.,
at about $\delta m^2\sim 2.5$--3.2 eV$^2$.}
\end{figure}

%---------------------------------------------------------------------------
\begin{figure}
\vspace*{0.6cm}\hspace*{-8mm}
\includegraphics[scale=0.86, bb= 100 100 500 720]{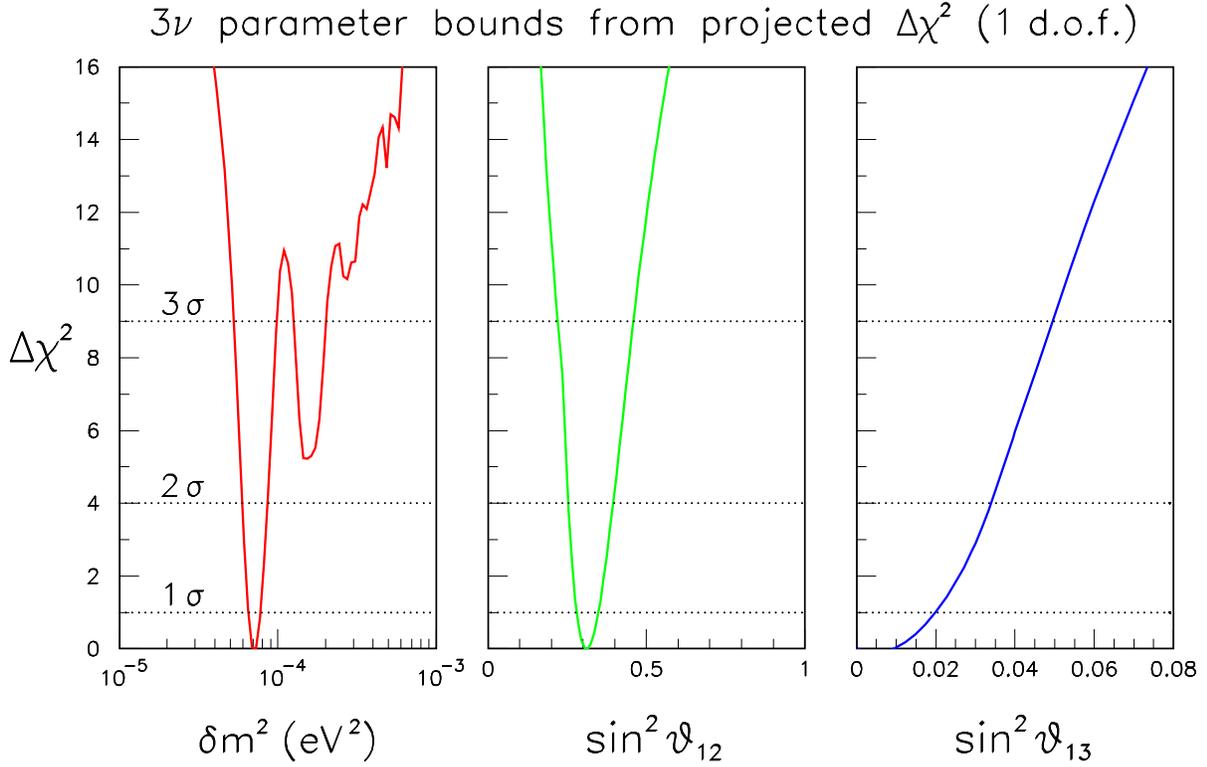}
\vspace*{-1.5cm} \caption{\label{fig9} Three-flavor active
neutrino oscillations: Projections of the global $(\Delta\chi^2)$
function onto each of the $(\delta
m^2,\sin^2\theta_{12},\sin^2\theta_{13})$ parameters. The
$n$-sigma bounds on each parameter (the others being
unconstrained) correspond to $\Delta\chi^2=n^2$. In the left
panel, the absolute minimum (LMA-I) and the second best fit
(LMA-II) are clearly visible. A third shallow minimum (LMA-III) of
the $\Delta\chi^2(\delta m^2)$ function is marginally present at
$\sim 3.2\sigma$.}
\end{figure}

%---------------------------------------------------------------------------
\begin{figure}
\vspace*{0.6cm}\hspace*{-8mm}
\includegraphics[scale=0.86, bb= 100 100 500 720]{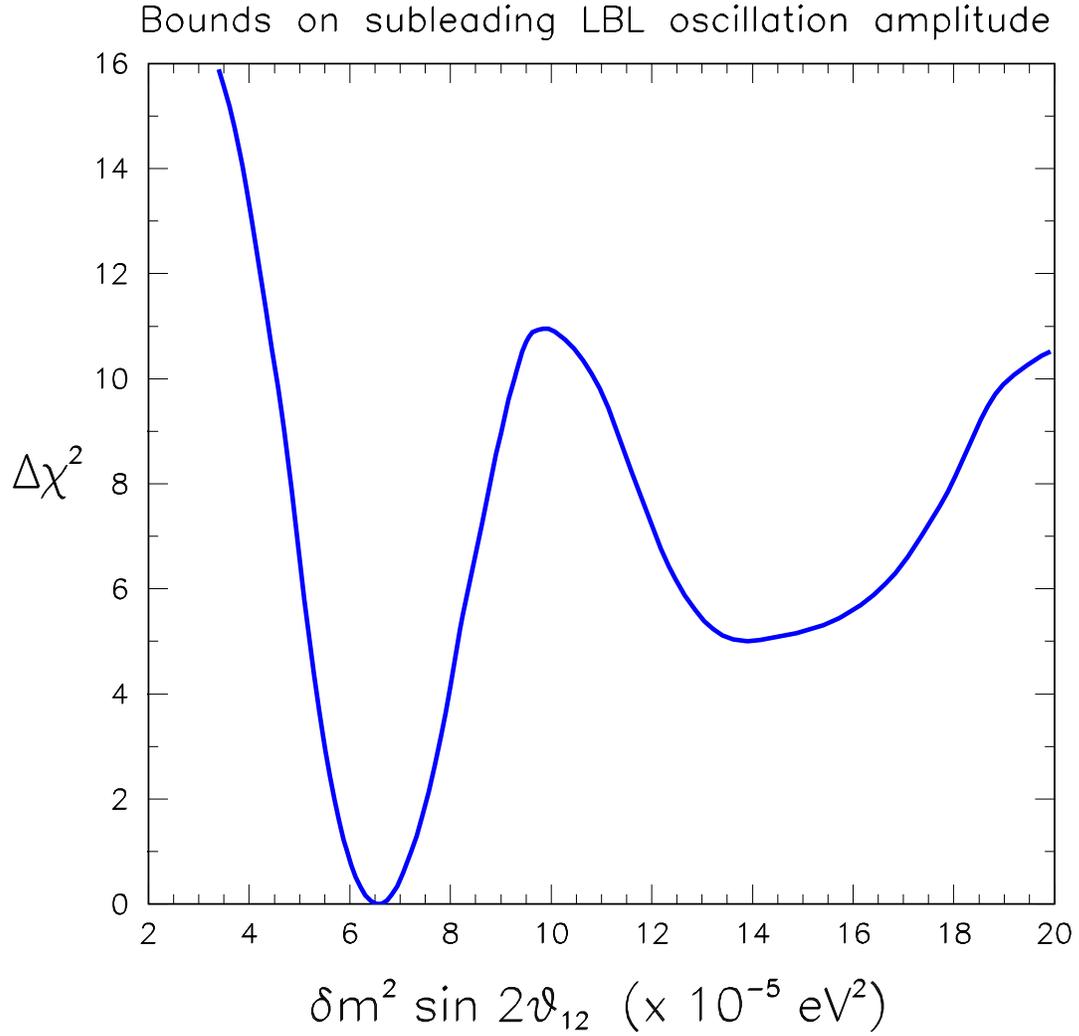}
\vspace*{-1.5cm} \caption{\label{fig10} Global bounds on the
parameter combination $\delta m^2\sin2\theta_{12}$, governing the
amplitude of subleading ``solar'' neutrino oscillations in long
baseline (LBL) experiments. See the text for details.}
\end{figure}

\end{document}